\documentclass{PoS}

\title{Direct-photon and heavy-flavour production in proton--proton collisions at $\sqrt{s} = 7$\,TeV with ALICE}

\ShortTitle{Direct-photon and heavy-flavour production in pp at $\sqrt{s} = 7$\,TeV with ALICE}

\author{Horst Sebastian Scheid
        for the ALICE Collaboration\\
        E-mail: \email{s.scheid@cern.ch}}

\abstract{
Low-mass dielectron measurements play an essential role in the study of the Quark-Gluon Plasma (QGP) created in ultra-relativistic heavy-ion collisions. They are produced in all stages of the collision and are not affected by final-state interactions. Thus, they provide a penetrating probe of the created medium. In the dielectron intermediate-mass region a measurement of the thermal radiation from the QGP gives information on the medium temperature. However, in this region the main component of the dielectron continuum stems from correlated semi-leptonic decays of charm and beauty hadrons. Therefore, it is crucial to understand the heavy-flavour (HF) production in vacuum and to find a way to separate this contribution from the thermal dielectron signal of the QGP. In this paper we will present the production of correlated $\rm e^+e^-$ pairs in pp collisions at $\sqrt{s} = 7$\,TeV with the expectations from known hadronic sources as a function of $m_{\rm ee}$, $p_{\rm T,ee}$, and the pair distance of closest approach to the primary vertex $\rm DCA_{ee}$. The extraction of the ratio of inclusive photons to decay photons and the charm and beauty cross sections from a fit of the data with different Monte-Carlo generators are presented, providing insight into the mechanisms of heavy-flavour production. In particular, we demonstrate how the $\rm DCA_{ee}$ variable will allow prompt and non-prompt dielectron pairs to be separated and thus help to disentangle the contribution of thermal radiation from the contribution of charm and beauty. Finally, we present the prospect of extending this analysis to proton--lead collisions.
}

\FullConference{International Conference on Hard and Electromagnetic Probes of High-Energy Nuclear Collisions\\
		30 September - 5 October 2018\\
		Aix-Les-Bains, Savoie, France}

\begin{document}

\section{Introduction}

ALICE is the dedicated experiment at the CERN LHC to study strongly interacting matter under extreme conditions, i.e. at high temperatures. These can be reached in ultra-relativistic collisions of heavy nuclei. In these collisions, a state of matter of matter that existed microseconds after the Big Bang is created: a plasma of quarks and gluons (QGP).
Dielectrons are considered a prime probe of this state of matter. Since they are produced in all stages of the collision and do not interact via the strong force, they carry information on the entire evolution of the system. Dielectrons are produced in the decays of pseudoscalar and vector mesons, the semi-leptonic decays of correlated hadrons carrying one charm or beauty quark, and from the internal conversion of direct photons.
With the creation of a hot system, additional sources are expected, i.e. thermal radiation from the QGP or hadron gas. The medium is thought to modify the properties of vector mesons due to partial chiral symmetry restoration, in particular the short-lived rho meson. 
In addition, the initial conditions of the collisions are expected to change compared to elementary collisions due to modifications of the parton distribution functions in nuclei. This can have a significant effect on the production cross section of pairs of heavy quarks. In the dielectron intermediate mass region (IMR, $1.1 < m_{\rm ee} < 2.7$\,GeV/$c^2$) a measurement of thermal radiation can give information on the temperature of the QGP.
This makes it crucial to first understand dielectron production in pp collisions, which provides a baseline for production in vacuum.  The effect of modified parton distributions in nuclei can be studied in proton-lead (p-Pb) collisions.  In this paper, the steps of the data analysis are explained and the first measurements of dielectron production in pp collisions at a centre-of-mass energy of $\sqrt{s} = 7$\,TeV~\cite{ref-ee} are presented along with the prospects for dielectron analysis in p--Pb collisions at $\sqrt{s_{\rm NN}} = 5.02$\,TeV.

\section{Data analysis and results}
The analysis is performed with pp data acquired by ALICE during the first data-taking period of the LHC in 2010. The integrated luminosity of the data sample is $L_{\rm int} = 6.0\pm0.2$\,nb$^{-1}$ and corresponds to 370 million minimum bias pp collisions.
Reconstructed electrons and positrons are are combined to create a spectrum of opposite-sign (OS) pairs. This includes not only the signal but also background, that can be purely combinatorial or have some residual correlation from jets or cross pairs from double Dalitz decays. The background is estimated by constructing a spectrum of same-sign (SS) pairs. 
Remaining acceptance differences for SS and OS pairs are approximated using event mixing and taken into account during the subtraction of the background. The spectrum is then corrected for tracking and particle identification inefficiencies within the ALICE acceptance ($p_{\rm T,e} > 0.2$\,GeV/$c$, $ \eta_{\rm e}<0.8 $).

The production cross section for dielectrons in pp collisions at $\sqrt{s} = 7$\,TeV as a function of the invariant mass of the pair ($m_{\rm ee}$) is compared to a so-called hadronic cocktail in the left panel of Fig. 1. The right panel shows the ratio of inclusive to decay photons ($R_{\gamma}$) extracted with a template fit in the mass region $0.09 < m_{\rm ee} <0.39$\,GeV/$c^2$. The procedure is described in detail in~\cite{ref-ee}.
\begin{figure}[ht]
\centering
  \begin{minipage}{0.47\textwidth}
    \includegraphics[scale=0.35]{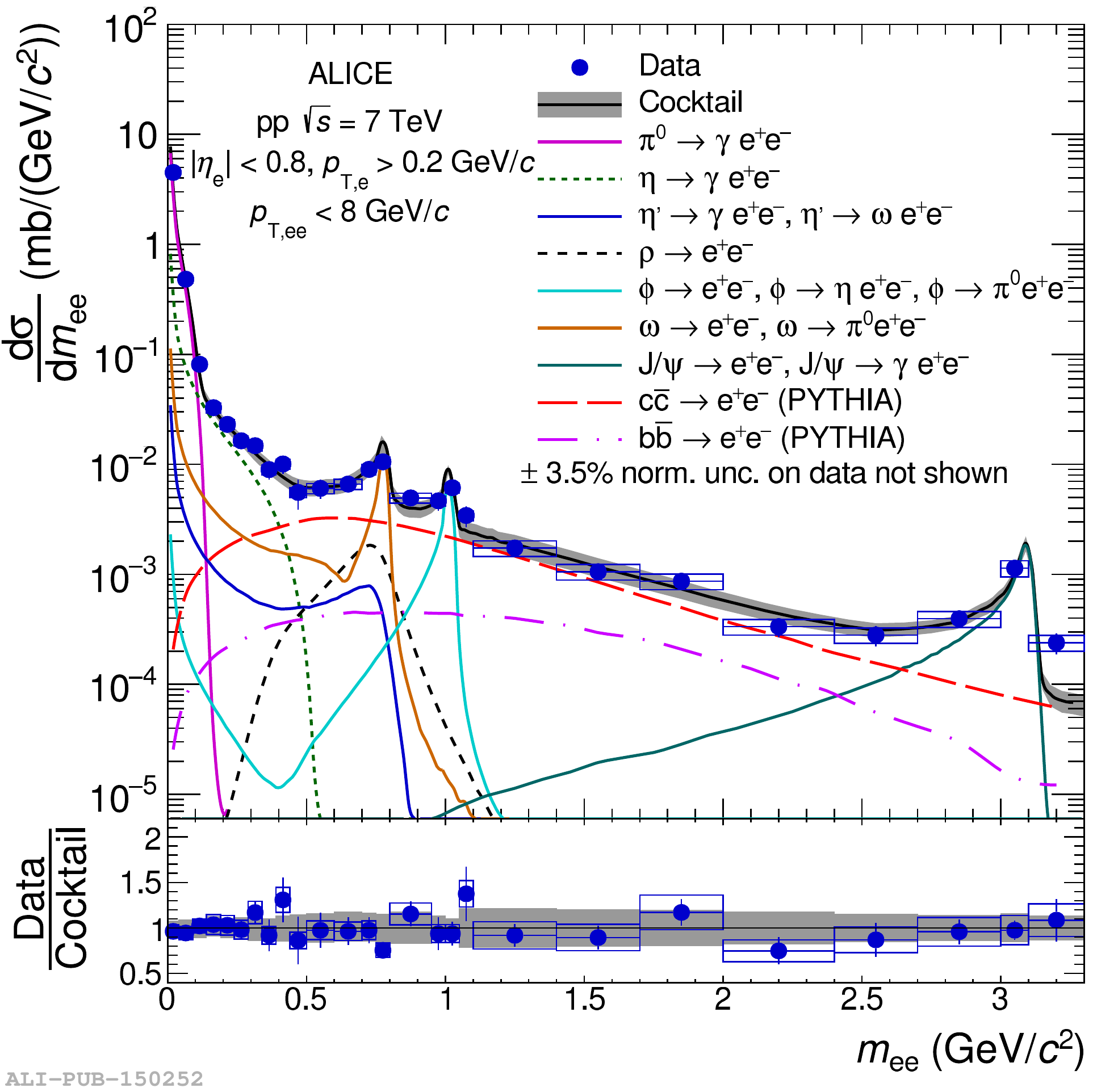}
  \end{minipage}
  \begin{minipage}{0.47\textwidth}
      \includegraphics[scale=0.35]{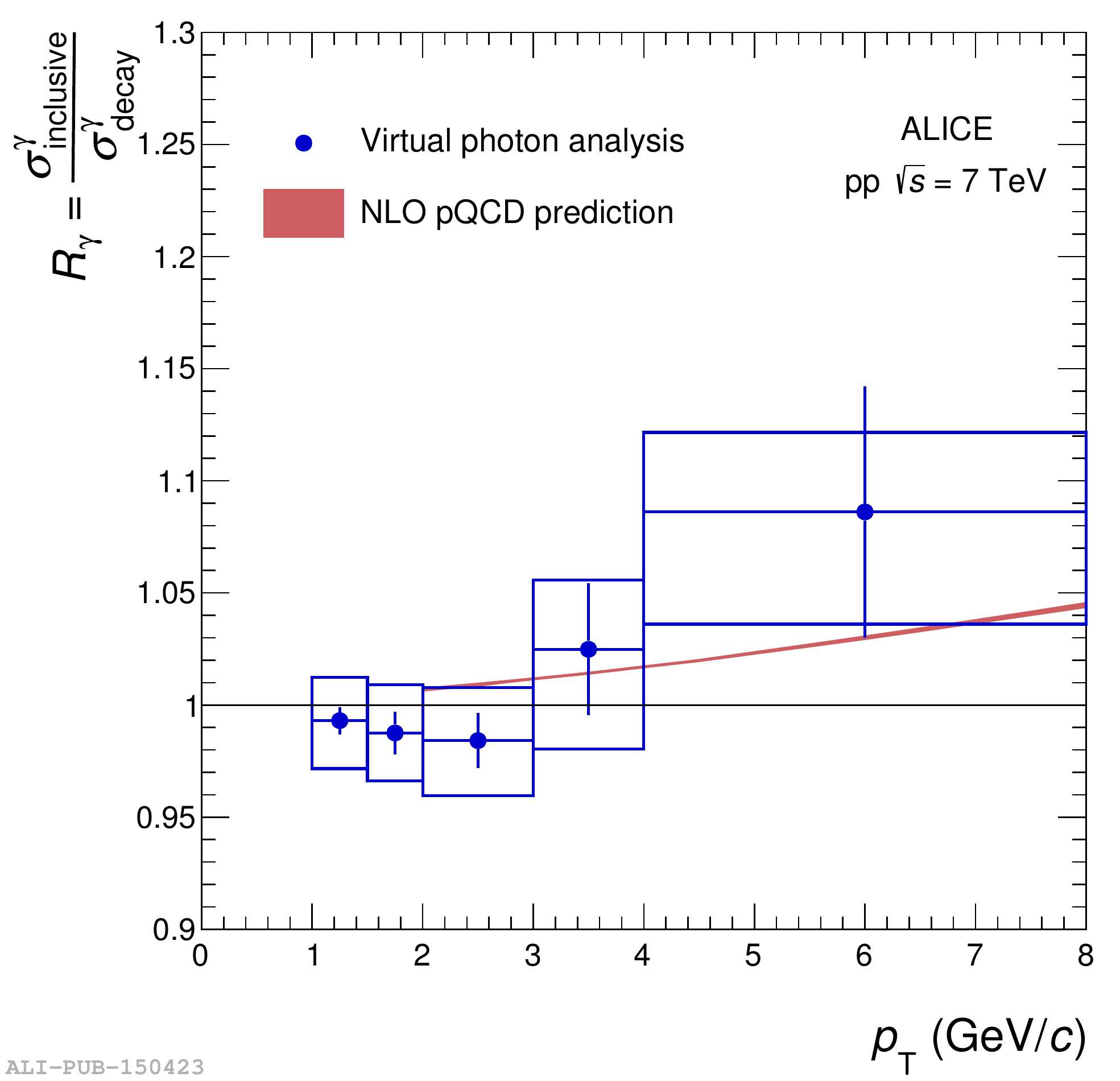}
  \end{minipage}
\caption{Dielectron cross section in pp collisions at $\sqrt{s} = 7$\,TeV as a function of $m_{\rm ee}$ (left) compared to a cocktail of known hadronic sources. Ratio of inclusive to decay photons ($R_{\gamma}$) (right) measured in pp collisions at $\sqrt{s}=7$\,TeV as a function of transverse momentum compared to NLO pQCD calculations.}
\end{figure}
The hadronic cocktail is an estimation of the contributions to the dielectron spectrum from known hadron decays.
The good agreement of the measured cross section with the expectation suggests a good understanding of dielectron production at LHC energies in the ALICE acceptance in pp collisions.

A detailed inspection of the dielectron yield at very low masses allows the dielectron photon production to be estimated by extrapolating the virtual photon yield using the Kroll-Wada equation. The ratio of inclusive photons to decay photons ($R_{\gamma}$), extrapolated from a fit in the mass region $0.09 < m_{\rm ee} < 0.39$\,GeV/$c^2$, is shown in the right panel of Fig. 1. Within large experimental uncertainties, the data are consistent with calculations from pQCD.

The main contribution in the IMR stems from decays of open heavy-flavour hadrons. The large relative cross section of $\rm c\bar{c}$ at LHC energies makes the measurement of thermal radiation in p--Pb and Pb--Pb collisions in this region especially  challenging. Experimental means to separate prompt thermal and non-prompt open heavy-flavour contributions to the dielectron spectrum have to be applied~\cite{ref-na60}. Our approach is to use the different decay kinematics of the two sources of dielectrons to separate them. The observable used is the pair-distance-of-closest-approach ($\rm DCA_{ee}$):
\begin{equation}
{\rm DCA_{ee}} = \sqrt{\frac{({\rm DCA_{{\it xy},1}}/\sigma_{xy{ \rm ,1}})^{2}+({\rm DCA_{{\it xy},2}}/\sigma_{xy,2})^{2}}{2}}.
\end{equation}
Here $\rm DCA_{\it xy,i}$ is the shortest distance of the track helix to the collision vertex in the transverse plane and $\sigma_{xy,i}$ its resolution. For a strong or electromagnetic decay, the electron track will always point to the primary vertex, whereas products of weak decays, i.e.\ decays with a finite decay length will not point to the vertex and thus produce larger $\rm DCA_{ee}$ values.

The dielectron cross section in the IMR is presented in Fig. 2 as a function of $\rm DCA_{ee}$ (left panel) and $p_{\rm T,ee}$ (right panel), respectively.
 \begin{figure}[ht]
\centering
  \begin{minipage}{0.47\textwidth}
    \centering
    \includegraphics[scale=0.35]{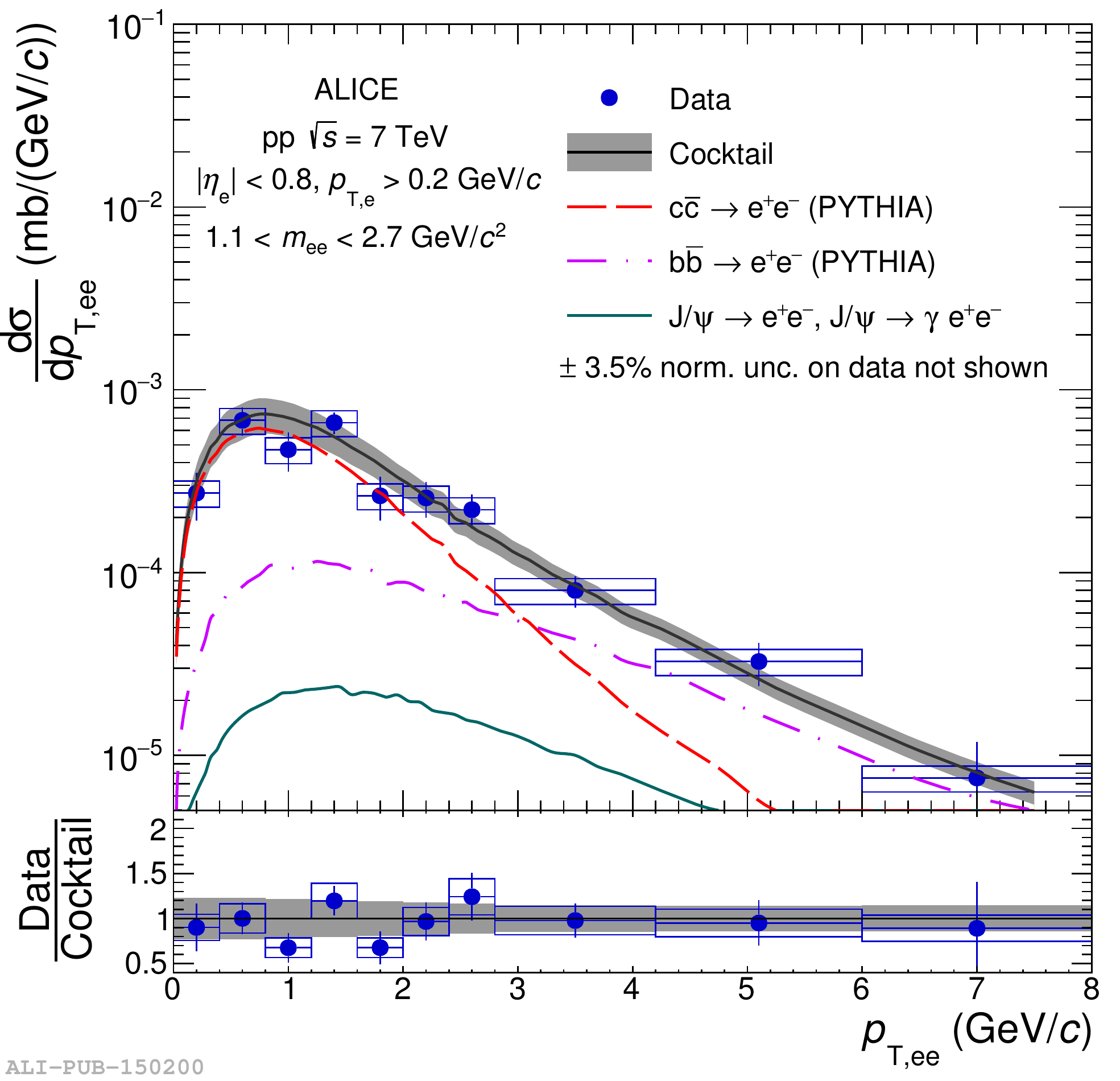}
  \end{minipage}
  \begin{minipage}{0.47\textwidth}
    \centering
    \includegraphics[scale=0.35]{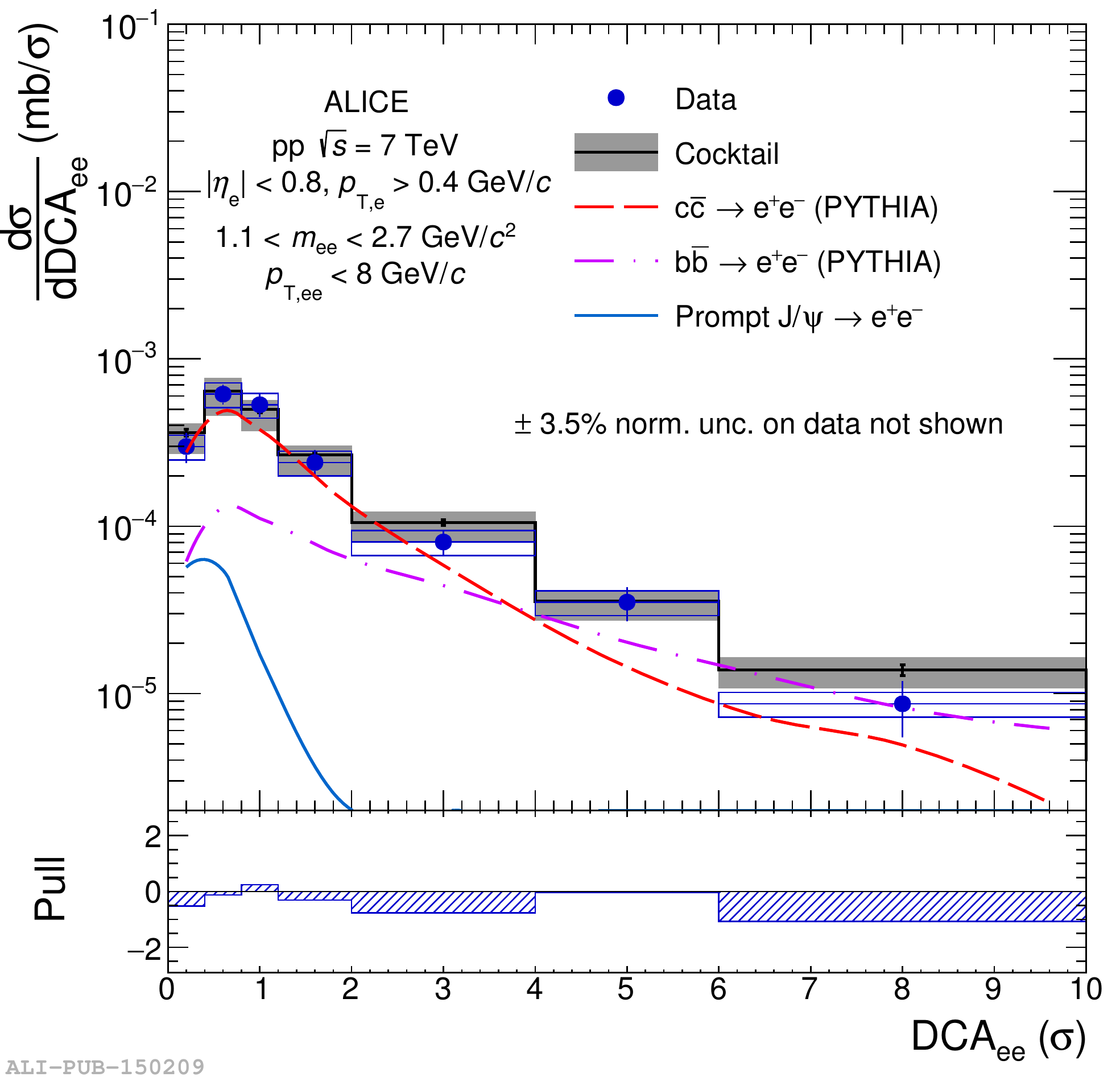}
  \end{minipage}
\caption{Dielectron cross section in pp collisions at $\sqrt{s} = 7$\,TeV as a function of $p_{\rm T,ee}$ (left) and $\rm DCA_{ee}$ (right) in the IMR compared to a cocktail calculated with PYTHIA~6~\cite{ref-ee}.}
\end{figure}
\begin{table}
  \centering
  \begin{tabular}[c]{cccc}
  && PYTHIA & POWHEG \\
  \hline
{$p_{\rm T,ee}/m_{\rm ee}$} & $\rm \sigma_{c\bar{c}}$& $\rm 6.4 \pm 0.9 (stat.) \pm 1.1 (syst.)$\,mb & $\rm 11.6 \pm 1.4 (stat.) \pm 1.9 (syst.)$\,mb \\

 					   & $\rm \sigma_{b\bar{b}}$& $\rm 0.303 \pm 0.077 (stat.) \pm 0.050 (syst.) $\,mb & $\rm 0.162 \pm 0.078 (stat.) \pm 0.026 (syst.) $\,mb \\ 
  \hline								
{$\rm DCA_{ee}$}                & $\rm \sigma_{c\bar{c}}$& $\rm 7.7 \pm 1.2 (stat.) \pm 1.3 (syst.)$\,mb & $\rm 11.7 \pm 1.8 (stat.) \pm 2.0 (syst.) $\,mb \\

 					   & $\rm \sigma_{b\bar{b}}$& $\rm 0.165 \pm 0.086 (stat.) \pm 0.028 (syst.) $\,mb & $\rm 0.175 \pm 0.092 (stat.) \pm 0.030 (syst.) $\,mb \\ 
 \end{tabular}
 \caption{Summary of the total $\rm c\bar{c}$ and $\rm b\bar{b}$ cross sections extracted from a fit of the measured dielectron spectra from heavy-flavour hadron decays in ($m_{\rm ee}$, $p_{\rm T,ee}$) and in $\rm DCA_{ee}$ with PYTHIA and POWHEG. The uncertainty of 22\% on the branching fractions and fragmentation functions ($\rm BR_{Q\rightarrow e}$) is not listed~\cite{ref-ee}.}
\end{table}
In this region the dielectron cross section is dominated by the contributions from open charm and beauty hadrons. The measured $\rm DCA_{ee}$ distribution shows no indication for a prompt dielectron source in this mass region.
To further investigate the production of heavy quarks we exchanged the leading-order event generator PYTHIA by the next-to-leading-order generator POWHEG~\cite{ref-powheg}. A two dimensional fit to $p_{\rm T,ee}/m_{\rm ee}$ and a fit of $\rm DCA_{ee}$  is used to extract the heavy-quark production cross section. The results of the $p_{\rm T,ee}/m_{\rm ee}$ and $\rm DCA_{ee}$ fits are summarised in Tab.~1 and are found to be consistent within the uncertainties and comparable with independent measurements from single HF particle spectra~\cite{ref-ccbar,ref-bbbar}. However, the uncertainties between the values obtained with PYTHIA and POWHEG are fully correlated and a strong model dependence can be observed, reflecting the sensitivity of the measurement to the heavy-flavour production mechanisms in the Monte-Carlo event generators.
\section{Prospects for the p--Pb analysis}
To single out the interesting characteristics of the QGP in heavy-ion collisions, it is necessary to understand the effects that the modification of the particle distribution functions have on the production of heavy quarks, and the possible presence of thermal radiation in small systems. This can be studied with high-statistics p--Pb data taken in 2016. 

\begin{figure}[ht]
\centering
  \begin{minipage}{0.47\textwidth}
    \includegraphics[trim={0, 0, 0, 1.5cm},clip,scale=0.35]{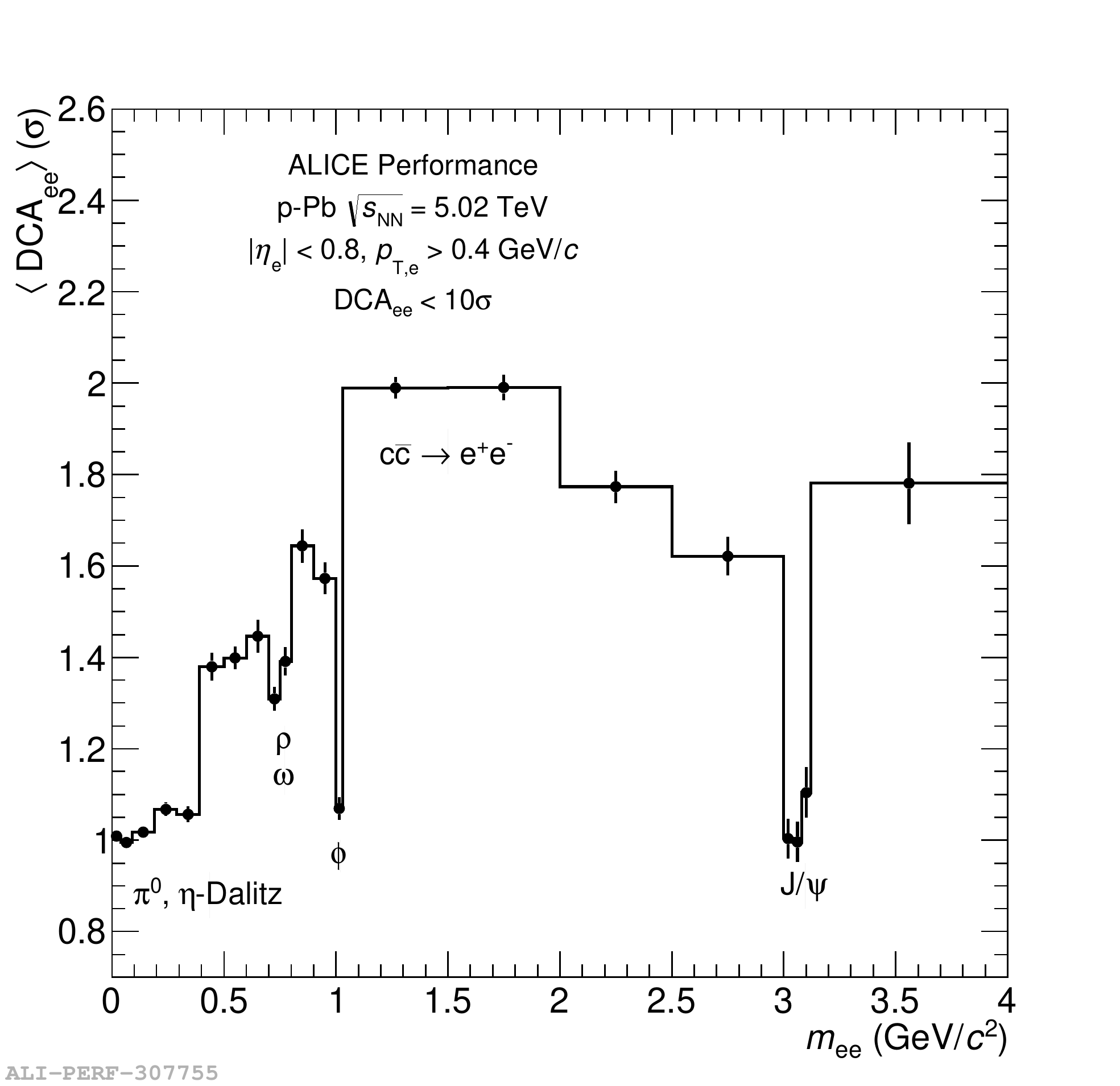}
  \end{minipage}
  \begin{minipage}{0.47\textwidth}
    \includegraphics[trim={0, 0, 0, 1.5cm},clip,scale=0.35]{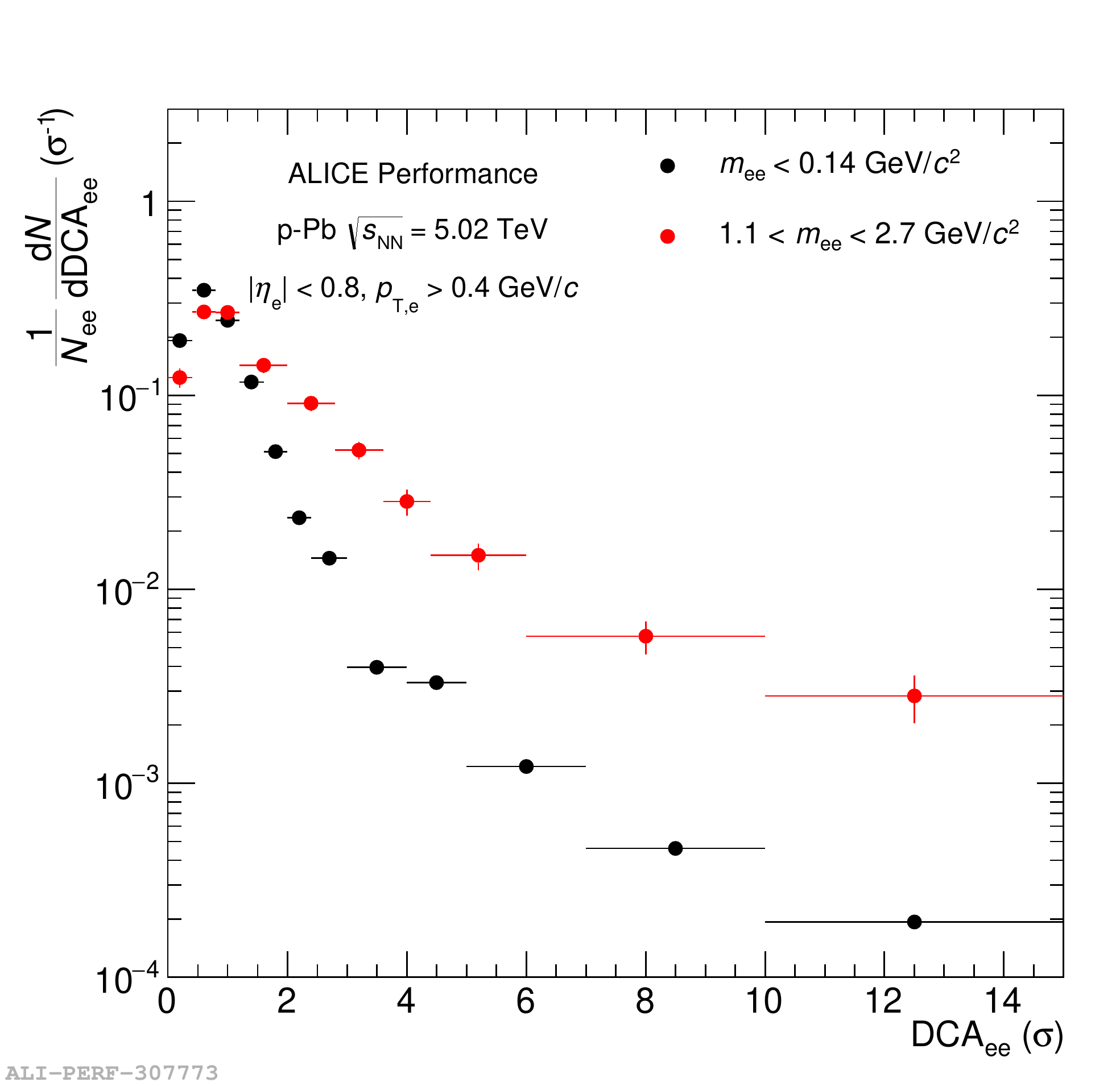}
  \end{minipage}
\caption{$\langle \rm DCA_{ee} \rangle$ as a function of $m_{\rm ee}$ (left) and $\rm DCA_{ee}$ spectra in the $\pi$-mass region and the IMR normalised to unity (right) in p-Pb collisions at $\sqrt{s_{\rm NN}} = 5.02$\,TeV.}
\end{figure}
First performance studies are presented in Fig. 3.
In the left panel, the $\langle \rm DCA_{ee} \rangle$ is shown in p--Pb collisions at $\sqrt{s_{\rm NN}} = 5.02$\,TeV as a function of $m_{\rm ee}$. We can clearly see a rise in $\langle \rm DCA_{ee} \rangle$ towards higher masses, where a saturation is reached for $m_{\rm ee} > m_{\rm \phi}$. Dips in the spectrum occur
at the masses of vector mesons. The observed structure can be understood as the relative abundance of open heavy-flavour contributions in the dielectron mass spectrum.
In the right panel of Fig. 3 we show the $\rm DCA_{ee}$ spectra for the mass region dominated by the $\pi^0$ compared to the one in the IMR. With the high statistics data it is possible to extend the reach of the analysis to $\rm DCA_{ee} < 15 \sigma$, which will give better constraints on the contribution of charm and beauty, a requirement to measure possible thermal radiation.
\section{Conclusion}
The measurement of the dielectron cross section in $\sqrt{s} = 7$\,TeV as function of $m_{\rm ee}$, $p_{\rm T,ee}$ and in $\rm DCA_{ee}$ is well described by the expectation from known hadronic sources, which implies a good understanding of the cross section of dielectrons in the ALICE acceptance. We show that $\rm DCA_{ee}$ gives the possibility to separate prompt and non-prompt sources and thus will play an important role in the determination of the temperature of the QGP and the study of thermal radiation in small systems at the LHC. The IMR can give further constraints on the production mechanisms of heavy-flavour quarks.


\begin{thebibliography}{999}
  \bibitem[1]{ref-ee}
    Acharya, S. {\em et al}, {\em JHEP} {\em 09}, {\bf 2018}, 64.
  \bibitem[2]{ref-pythia6}
    Sjostrand, T. and Mrenna, S. and Skands, P., {\em JHEP} {\em 06}, {\bf 2006}, 26.
  \bibitem[3]{ref-wa80}
    L. Altenk{\"a}mper, F. Bock, C. Loizides, and N. Schmidt, {\em Phys. Rev. C 96}, {\bf 2017}, 064907.
  \bibitem[4]{ref-perugia2011}
    Skands, P., {\em Phys. Rev.} {\em D82}, {\bf 2010}, 26.
  \bibitem[5]{ref-ccbar}
        Acharya, S. {\em et al}, {\em Eur. Phys. J.} {\em C77}, {\bf 2017}, 550.
  \bibitem[6]{ref-bbbar}
    Aaij, R. {\em et al}, {\em Eur. Phys. J.} {\em C71}, {\bf 2011}, 1645.
  \bibitem[6]{ref-na60}
    Arnaldi, R. {\em et al}, {\em Eur. Phys. J.} {\em C59}, {\bf 2009}, 607.
  \bibitem[7]{ref-powheg}
    Nason, P., {\em JHEP} {\em 11}, {\bf 2004}, 40.
  
\end{thebibliography}
\end{document}